# An examination of local strain fields evolution in ductile cast iron through micromechanical simulations based on 3D imaging

Victor Manuel Trejo Navas[1], Ante Buljac[2,3], François Hild[2], Thilo F. Morgeneyer[3], Marc Bernacki[1], and Pierre-Olivier Bouchard[1]

[1] MINES ParisTech, PSL University, CEMEF-Centre de mise en forme des matériaux, CNRS UMR 7635, Sophia Antipolis, France
[2] Université Paris-Saclay, CentraleSupélec, ENS Paris-Saclay, CNRS, LMPS - Laboratoire de Mécanique Paris-Saclay, Gif-sur-Yvette, France
[3] MINES ParisTech, PSL Research University, Centre des Matériaux, CNRS UMR 7633, Evry, France

Microscopic digital volume correlation (DVC) and finite element precoalescence strain evaluations are compared for two nodular cast iron specimens. Displacement fields from *in-situ* 3D synchrotron laminography images are obtained by DVC. Subsequently the microstructure is explicitly meshed from the images considering nodules as voids. Boundary conditions are applied from the DVC measurement. Image segmentation-related uncertainties are taken into account and observed to be negligible with respect to the differences between strain levels. Macroscopic as well as local strain levels in coalescing ligaments between voids nucleated at large graphite nodules are compared. Macroscopic strain levels are consistently predicted. A very good agreement is observed for one of the specimens, while the strain levels for the second specimen presents some discrepancies. Limitations of the modeling and numerical framework are discussed in light of these differences. A study on the use of strain as coalescence indicator is initiated.

## 1 Introduction

A better understanding of structure-property relationships may lead to the development of improved materials. In the case of ductile materials, efforts are put on the knowledge and modeling of the micromechanisms of ductile damage, namely, void nucleation, growth and coalescence. A very common limitation of fracture criteria is that they are validated for a single and usually proportional loading condition (Papasidero et al. 2015; Tekkaya et al. 2020). More versatile and accurate fracture criteria are needed. A sound basis for the micromechanics of damage would constitute an important benefit for the formulation of more universal fracture criteria and improved macroscopic constitutive laws. Numerous efforts have been dedicated to the analysis (Weck and Wilkinson 2008; Buljac et al. 2018a) and modeling (Scheyvaerts et al. 2010) of void nucleation and void growth (Koplik and Needleman 1988; Huang 1991; Kuna and Sun 1996). Coalescence remains, however, a complex and not completely understood phenomenon in which loading conditions, plastic flow characteristics and different microstructural factors play an important role (Pineau et al. 2016).

Three-dimensional imaging techniques such as tomography (Gammage et al. 2005; Weck et al. 2008) and laminography (Helfen et al. 2005; Shen et al. 2013) provide highly detailed and promising experimental data on the understanding of ductile damage all the way up to fracture (Buljac et al. 2016; Buljac et al. 2018a). This work constitutes a study of void coalescence through the examination of microscopic precoalescence strain measurements in ductile cast iron thanks to simulations based on 3D imaging.

Previous studies have investigated microscopic strain measurements in cast iron. Kasvayee et al. (2016) carried out a comparison between strain measurements predicted by finite element (FE) simulations and local strain fields measured via digital image correlation (DIC). Salomonsson and Olofsson (2017) and Seleš et al. (2019) studied strain localization via 3D FE simulations of heterogeneous cast iron microstructures and highlighted the three dimensional complexity of the strain field as the possible reason for disagreement between local predictions of strains and digital volume correlation (DVC) measurements. Buljac et al. (2018a) used bulk strain measurements obtained via DVC to study void coalescence mechanisms in nodular graphite cast iron.

Critical values of local strains have been investigated as a microscopic coalescence indicator for internal necking (Weck and Wilkinson 2008; Buljac et al. 2018a), void-sheet or shear band formation (Bandstra and Koss 2001; Weck and Wilkinson 2008; Buljac et al. 2018a) and fatigue crack initiation (Fischer et al. 2013). These critical strain values are based on cumbersome experimental observations and DVC analyses. Their accuracy relies on the size of the DVC probe, which may sometimes be larger than the ligament in particular at the onset of coalescence. In addition, the DVC approach presented in (Buljac et al. 2018a) is restricted to the identification of a critical strain prior to coalescence in the ligament. This critical strain can be analyzed with respect to the type of coalescence mechanism observed experimentally. But it cannot be used together with the local stress loading path, which could have an impact on this critical local strain to fracture. A numerical approach based on FE computations would therefore be interesting since it would enable the accuracy of the strain field to be checked prior to coalescence in addition to the influence of the probe size (DVC or FE). Once validated, it can be used to perform such strain to coalescence analyses in the whole microstructure and this would therefore enable a large range of stress states to be accounted for. The relationship between local strain to coalescence as a function of stress loading path could be studied. However, FE computations need first to be validated in terms of accuracy. This requires the ability for the FE approach to represent as precisely as possible the real microstructure and the boundary conditions applied to the region of interest. The goal of this paper is to compare strain values computed by FE simulations and evaluated by DVC in order to validate the numerical approach.

In this work, precoalescence strains are assessed with a numerical-experimental framework that combines laminography images acquired during mechanical tests, displacement and strain measurements via DVC and 3D FE simulations with immersed microstructures and realistic boundary conditions. The novelty of this work lies in the ability to analyze local strain prior to coalescence for large plastic strains, with extreme FE geometrical discretization, in 3D and with exact boundary conditions. The methodology is described in Section 2. In Section 3, predicted FE strains in four different probe volumes are then validated with DVC measurements via a thorough comparison after investigating the numerical sensitivity to mesh and probe sizes. The change of the strain fields is assessed in detail and the suitability of strain as a coalescence indicator is discussed.

## 2 Methodology

The general methodology of this work has been extensively documented in (Buljac et al. 2017; Shakoor et al. 2017c). Consequently, only the key aspects are briefly recalled hereafter to provide the necessary background. The core of the methodology is a combination of Synchrotron Radiation Computed Laminography (SRCL), DVC and FE simulations. Figure 1 shows the two tensile specimens studied where macroscopic holes were machined respectively at 45° and 90° with respect to the loading direction. This enables the macroscopic stress to be varied in the region of interest since it is known to be an important factor for ductile damage.

### 2.1 Laminography

SRCL is a non-destructive technique that produces three-dimensional images of a scanned region (Helfen et al. 2005). The difference between tomography and laminography is that the latter is applied to laterally extended three-dimensional specimens as opposed to the stick-like geometry associated with tomography.

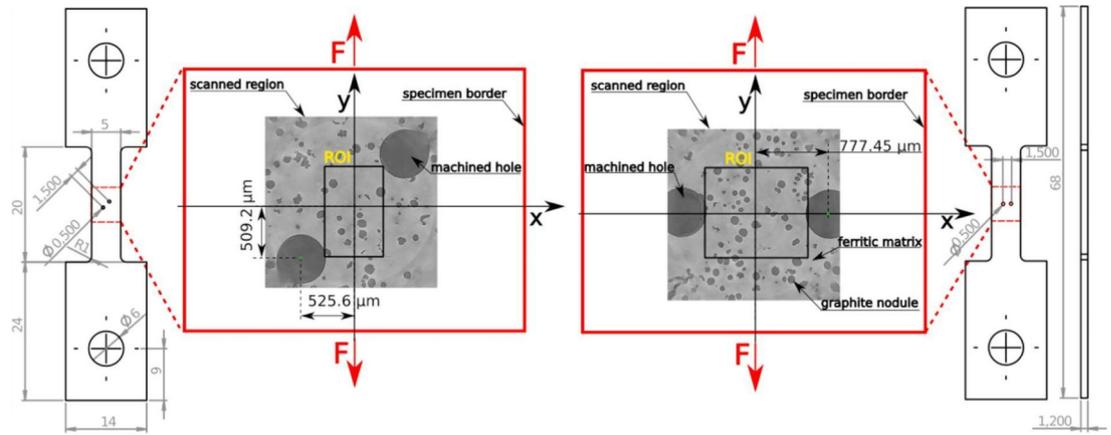

**Figure 1** Geometry of the two studied specimens: 45° on the left and 90° on the right. The zone scanned with SRCL and the region of interest (ROI) used in DVC calculations are indicated. The loading direction is depicted with red arrows. Adapted after (Buljac et al. 2018a).

*In-situ* mechanical tests are carried out to monitor the three-dimensional evolution of the microstructure all the way up to fracture (Buljac et al. 2018a). Displacements are applied with a manual device via screw rotation. The applied displacements are recalled in Table 1. A reconstructed volume of size 1600 × 1600 × 1600 voxels is obtained after each scan. The side length of one cubic voxel is 1.1 µm (Buljac et al. 2018a).

**Table 1** Nominal cross head displacement applied to each of the studied specimens.

|  | Cross head displacement [µm] | |
| --- | --- | --- |
| Increment | 45° specimen | 90° specimen |
| 0 | 0 | 0 |
| 1 | 248 | 186 |
| 2 | 372 | 248 |
| 3 | 496 | 289 |
| 4 | 620 | 330 |
| 5 | 744 | 372 |
| 6 | 868 | 413 |
| 7 | 992 | 454 |
| 8 | 1074 | 496 |
| 9 | 1136 | 537 |
| 10 | 1198 | 578 |
| 11 | 1260 | 599 |
| 12 | – | 620 |
| 13 | – | 641 |
| 14 | – | 661 |
| 15 | – | 682 |
| 16 | – | 703 |
| 17 | – | 722 |
| 18 | – | 744 |
| 19 | – | 765 |

### 2.2 Digital volume correlation

Digital Volume Correlation (DVC) is a technique that allows three-dimensional displacement fields to be measured (Bay et al. 1999) based on the gray level conservation of two SRCL images. The technique consists in finding the displacement field $u(x)$ that globally minimizes (Roux et al. 2008) the L2-norm of gray level residuals $\rho(x)$ between the reference configuration $f$ and the deformed configuration $g$:

$$\rho(x) = f(x) - g[x + u(x)]. \tag{1}$$

The laminography images are used as input for the DVC procedure. To reduce the computational cost, blocks of 8 voxels called supervoxels are used with the corresponding mean gray level. The used DVC element side length is 16 supervoxels. This yields a standard equivalent strain uncertainty of 0.78 % for the 90° specimen and 0.73 % for the 45° specimen. Further details are given in (Buljac et al. 2018b). It is the microstructural heterogeneities that provide the necessary gray level contrast. The resulting displacement fields can be used to calculate the three-dimensional strain field in the region of interest (ROI).

### 2.3 Numerical simulations

Numerical simulations are conducted using a Lagrangian FE P1+/P1 formulation (Brezzi et al. 2008) in the CimLib library (Digonnet et al. 2007). Monolithic multiphase simulations are carried out with this formulation at the microscale (Shakoor et al. 2015). An important ingredient in the numerical framework is the use of body-fitted meshes at the interfaces since it ensures excellent volume conservation of the phases (Shakoor et al. 2017b). The FE mesh is adapted with respect to the maximum principal curvature $\lambda_{max}$ (Quan et al. 2014; Shakoor et al. 2017a). The isotropic metric $M = \mathrm{diag}(1/h^2)$ is selected, where $h$ is the mesh size.

The graphite nodules provide a very low load-bearing capacity (Hütter et al. 2015; Tomičević et al. 2016). For this reason, their mechanical contribution is considered negligible in this work, i.e., they are considered as voids. The void phase is described as a Newtonian fluid with viscosity $\eta = 2.1$ MPa/s. An elastoplastic behavior $\sigma_{eq} = \sigma_y + K\varepsilon_{pl}^n$ is assigned to the ferritic matrix with Ludwik's isotropic hardening law (Ludwik 1909), where $\sigma_{eq}$ is the von Mises equivalent stress, $\varepsilon_{pl}$, the accumulated equivalent plastic strain, $\sigma_y$ the yield stress, $K$ the hardening modulus and $n$ the hardening exponent. A Young modulus of 210 GPa and a Poisson ratio $\nu = 0.3$ are selected. A calibration procedure at the microscale by means of X-ray microtromography, DVC and FE simulations (Buljac et al. 2018c) allowed the values of $\sigma_y$, $K$ and $n$ to be determined. The identified values were $\sigma_y = 245$ MPa, $K = 330$ MPa and $n = 0.21$.

The size of the simulation domain is $456 \times 818 \times 900\ \mu m^3$ for the 45° specimen and $1100 \times 340 \times 878\ \mu m^3$ for the 90° specimen.

### 2.4 SRCL-DVC-FE framework

In this work, FE simulations are carried out within an experimental-numerical framework (Buljac et al. 2017; Shakoor et al. 2017c) that allows simulations with immersed microstructures and realistic boundary conditions to be conducted. The three pillars of this methodology are i) laminography images acquired during *in-situ* mechanical tests, ii) three-dimensional measurement of displacement fields via DVC applied to the acquired images, and iii) FE simulations of the immersed microstructures that use the measured displacement fields as boundary conditions. All FE simulations presented herein are carried out within the SRCL-DVC-FE framework, but will simply be referred to as FE simulations for the sake of brevity. A schematic flowchart of the SRCL-DVC-FE framework is provided in Figure 2.

**Figure 2**  Flowchart of the SRCL-DVC-FE framework.

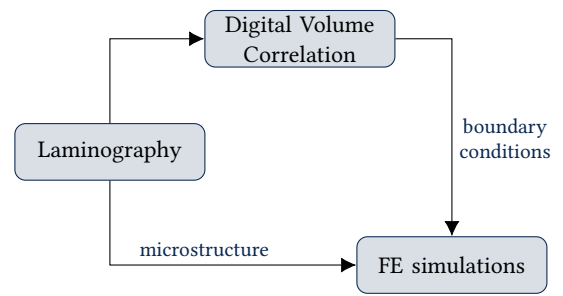

Despite monotonically increasing macroscopic loading conditions, local unloading may occur at the microscale due to:
- Particles/matrix debonding: for this material, the interface between the matrix and the nodules is considered as already debonded.
- Nucleation at a lower scale: this mechanism is not taken into account here since it is assumed that the dominant mechanism is void growth and coalescence.
- Coalescence: local probes are placed at the intervoid ligaments where coalescence first occurs. In addition, the FE model is built in such a way that coalescence by internal necking is accounted for, which means that the numerical model also accounts for such local unloading.

It is therefore considered here that the local total strain measured/computed in the ligament prior to fracture corresponds to the coalescence strain to fracture. Since von Mises plasticity is used here, this work does not account for possible local softening in the matrix that would be due to the presence of a second void population at a lower scale.

### 2.5 Observables

The observables selected for each specimen are:
- the local volume-averaged strain evaluations at two different locations corresponding to the first occurrence of coalescence, and
- the global (or macroscopic) volume-averaged strains over the whole simulation domain.

The local strain measurement corresponds to the cumulated equivalent von Mises strain in cubic probes of the size of one DVC element (unless otherwise stated). The shape and size of the FE probes were chosen in order to facilitate a direct comparison with DVC results. The simulated microstructural region is shown in Figure 3 for the 90° and 45° specimens. Probe 1 of each specimen has previously been associated with the void-sheet coalescence mechanism and probe 2 with internal necking (Buljac et al. 2018a). A particularity of the ligament associated with probe 2 of the 90° specimen is that it is not large enough to entirely contain the DVC probe.

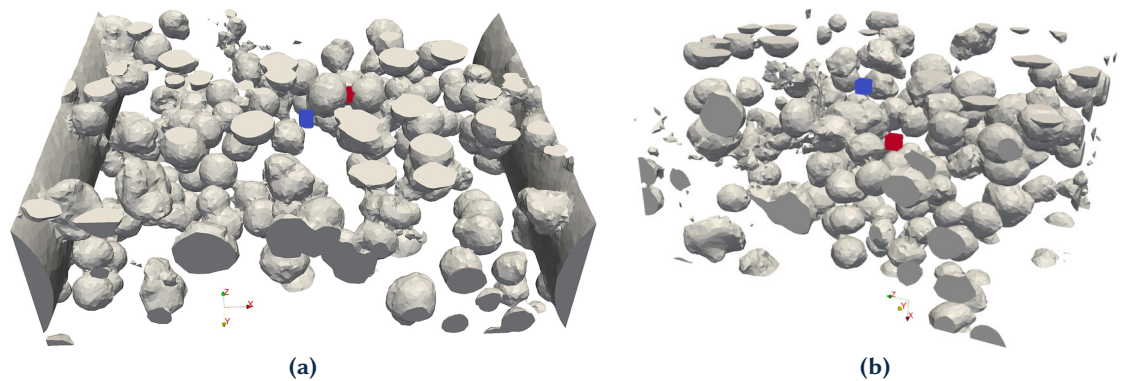

(a)     (b)

**Figure 3**  3D view of the simulated microstructure for (a) the 90° specimen and (b) the 45° specimen. Probed volumes are shown in blue (probe 1) and red (probe 2).

## 3 Results and discussion

### 3.1 Comparison of DVC and FE strain levels

In this section, a comparison of macroscopic and local strain measurements obtained from DVC and FE calculations is presented for both specimens. An estimation of the segmentation-related uncertainty was previously carried out (Trejo Navas et al. 2019), and the FE strain measurements presented in this section are the result of this procedure. The latter consisted of carrying out three different FE simulations, each with an immersed microstructure originating from a segmented image with a different method. Each FE strain measurement thus corresponds to the average of these three strain evaluation and the corresponding standard deviation is reported as the estimated uncertainty.

#### 3.1.1 90° specimen

The comparison of DVC and FE results in terms of the macroscopic equivalent strain of the 90° specimen along with the absolute difference between the two datasets is shown in Figure 4. A very good agreement is observed up to increment 10. Afterward, the DVC results are overestimated by the FE simulations with an increasing difference that reached 2 % strain at the final increment. This type of agreement for the macroscopic strain was expected since the FE simulations were driven by displacements obtained by means of DVC.

A comparison of the DVC and FE microscopic strain levels in probe 1 of the 90° specimen is shown in Figure 5(a). A very good agreement is observed between the two quantities except for the last increment where the DVC strain experiences an accelerated increase that the FE simulation does not capture. Figure 5(a) also presents the contours of the void phase obtained from the FE simulation superimposed with the corresponding slice of the laminography images for three different increments. The position of probe 1 is indicated by the blue box.

**Figure 4** Comparison of DVC and FE macroscopic equivalent strain measures for the 90° specimen. The error bars represent the segmentation-related uncertainty.

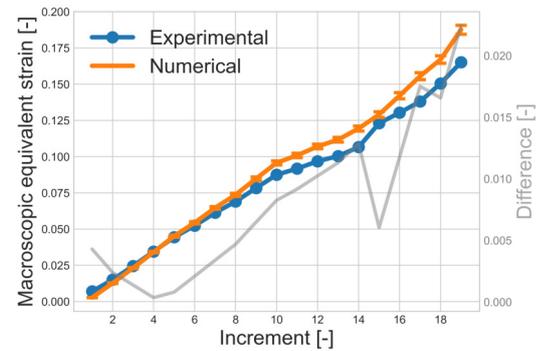

DVC and FE strain levels in probe 2 of the 90° specimen are reported in Figure 5(b). Although coalescence occurs at increment 11 by internal necking, the results are reported up to the end of the simulation for the sake of completeness. It is made possible with the FE results because the probe deforms with the matrix, and during coalescence, it flows out of the disappearing intervoid ligament. The measured strain in probe 2 is in this sense no longer associated with the intervoid ligament after increment 11. The DVC and FE results in probe 2 exhibit a very good correspondence up to increment 16 with an approximately constant strain difference of the order of 0.01. In the subsequent increments and up to the end of the simulation, the DVC values increase at a much higher rate than the FE values. A final and maximum strain difference of 0.20 is obtained.

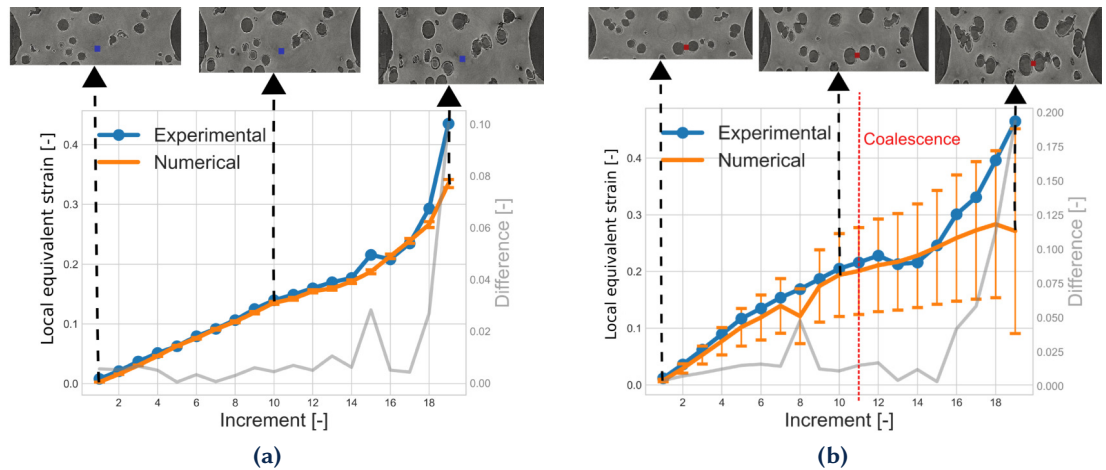

**Figure 5** Comparison of microscopic DVC and FE equivalent strain measures of the 90° specimen in (a) probe 1 and (b) probe 2. For three different increments, a slice of the original image is superimposed with the interfaces according to the FE simulation (white lines). Position marked with blue box for probe 1 and red box for probe 2. Error bars represent the segmentation-related uncertainty.

In the 90° specimen simulation, the amplitude of the reported segmentation-related uncertainty for the macroscopic strains and those in probe 1 is negligible with respect to the magnitude of the difference between the DVC and FE estimates. However, for the strains in probe 2, the uncertainty increases considerably during the simulation and is of the same order of magnitude as the strain levels. The reason for this difference between probes 1 and 2 is that the intervoid ligament in which probe 2 is located is not big enough to accommodate the whole DVC probe. This feature makes the results for probe 2 very sensitive to the position of the void-matrix interface.

As evidenced in Figure 5, void growth and shape changes are generally described satisfactorily. There are, however, some minor discrepancies in the position of the interfaces between the FE simulations and the laminography images. In some instances, void growth is underestimated and in others, overestimated. The magnitude of these discrepancies increases during the simulation.

### 3.1.2   45° specimen

In the case of the 45° specimen, see Figure 6, the FE macroscopic equivalent strain slightly overestimates the DVC levels for the first two increments. Then the two values coincide remarkably well up to increment 7 where the DVC strain start to accelerate, which is not well reproduced by the FE simulation. At the end of the simulation, a maximum strain difference of 0.04 is observed. This level is still acceptable given the corresponding strain (i.e., 0.25).

**Figure 6**   Comparison of DVC and FE macroscopic equivalent strain for the 45° specimen. Error bars represent the segmentation-related uncertainty.

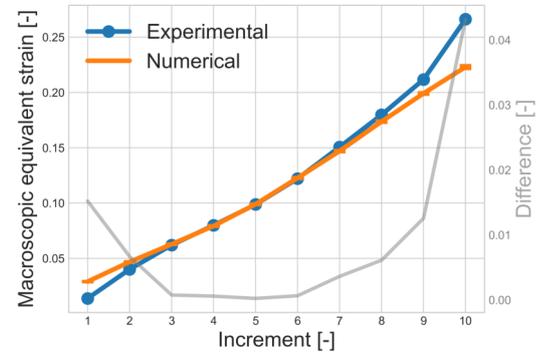

Although the FE simulation of the 45° specimen reproduces the rate of strain at the local probes, it considerably overestimates the DVC values for probe 1 (Figure 7(a)) and underestimates the DVC strain levels for probe 2 (Figure 7(b)). Similarly to the macroscopic equivalent strain, the DVC measured values in the probes show an accelerated increase toward the end of the test, which is not captured by the FE simulation. A maximum strain difference of 0.08 is observed for probe 1, and of 0.10 for probe 2. A possible reason for the offset between local DVC and FE results in the 45° specimen is the uncertainty with respect to the position of the DVC element. Although the size and shape of the FE probe was chosen to match those of the DVC element, the exact positions of the DVC element and of the probed volume may not coincide exactly. In order to place the FE probes, it was assumed that the coordinates probed for the DVC strain corresponded to the center of a DVC element. The uncertainty in the relative position of the probed volumes with respect to the DVC elements could introduce a non negligible strain uncertainty. This is an aspect to be improved in future validation works.

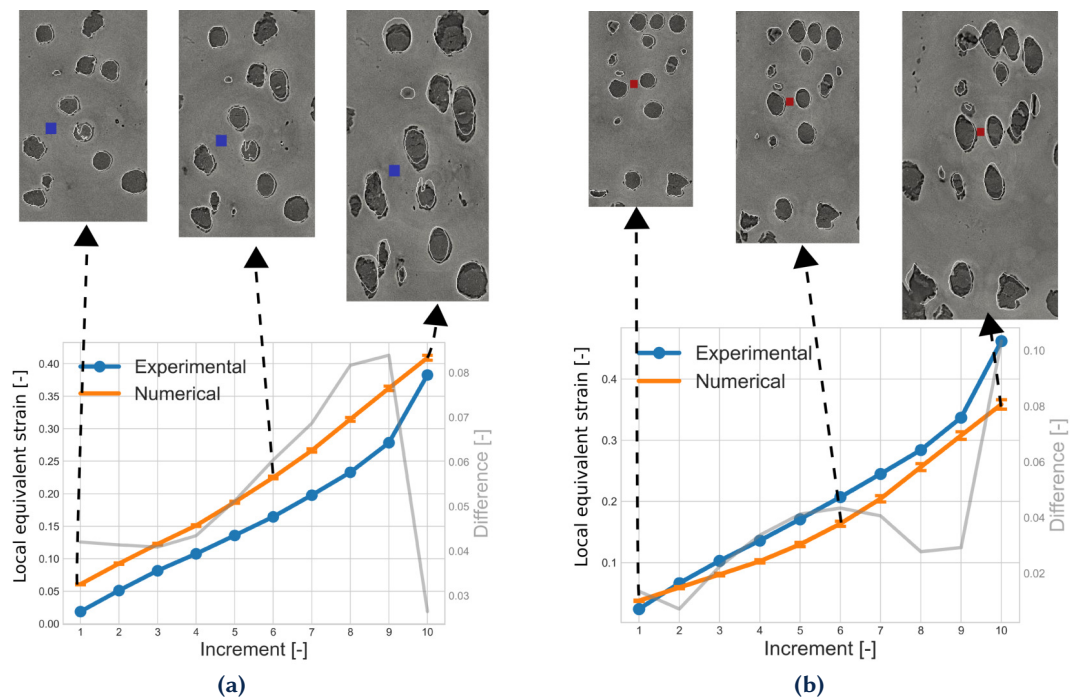

**Figure 7**   Comparison of DVC and FE equivalent strain evaluations of the 45° specimen in (a) probe 1 and (b) probe 2. For three different increments, a slice of the original image is superimposed with the interfaces according to the FE simulation (white lines). Position marked with blue box for probe 1 and red box for probe 2. Error bars represent the segmentation-related uncertainty.

Similar to the 90° case, in the shown slices (Figure 7) of the 45° case, albeit some local differences, an overall good agreement between the predicted position of the interfaces by the FE simulations, and the laminography images is observed. The differences increase toward the end of the simulation. The magnitude of the segmentation-related uncertainty is significantly smaller than the difference between the DVC and FE results for the macroscopic equivalent strain of the 45° specimen and the local strain for the two probes. This uncertainty was significant only for the strains in probe 2 for the aforementioned reason, see Section 3.1.2. The influence of the segmentation procedure on the results of the simulations has been previously discussed (Trejo Navas et al. 2019). The comparison with the DVC results presented in this section while taking into account this uncertainty leads to the conclusion that the differences between the DVC and the FE results are attributable to current limitations in the numerical framework and can be used as indications to improve the current framework.

One possible limitation of the mechanical model is pointed out by the inability of the FE simulations to reproduce the strain acceleration toward the end of the simulation for both specimens. This trend is observed in spite of efforts to calibrate the elastoplastic material parameters at the microscale via an experimental-computational identification framework with X-ray microtomography (Buljac et al. 2018c). This difference may point out a deficiency of the employed hardening law, namely, excessive hardening for large strain levels. A solution can be the use of hardening laws that saturate for large values of strain such as the Voce postulate (Voce 1955). This result is consistent with previous observations via *post mortem* Scanning Electron Microscopy (SEM) fractography of a smaller void population that is not visible in the laminography images due to their size with respect to the spatial resolution (2.19 μm per voxel). These smaller voids are neither explicitly modeled in the FE mesh nor implicitly considered in the constitutive law. Accounting for softening induced by the smaller void population is a perspective to this work. This could be achieved for example through the use of a coupled damage model in the matrix (El Khaoulani and Bouchard 2013).

Despite the differences in local strain levels between DVC measurements and FE predictions in the 45° case, the results are considered satisfactory. A very good agreement was obtained for the macroscopic strain of both specimens and for the precoalescence local strains of the probes of the 90° specimen. The local strain levels in the probes of the 45° specimen were not predicted accurately, but their rate of increase was well reproduced. The FE strain predictions are thus considered satisfactory and can now be used for the study of void coalescence.

## 3.2 Strain field history

The strain results provided by the FE simulations were validated with comparison with DVC measurements in Section 3.1 and their limitations as well as future aspects to be improved were discussed. This validation allows the FE results to be exploited with reasonable confidence in the study of void coalescence. To start this study, the change of the strain field in given slices of the domain is studied.

Figure 8 presents the change of the equivalent strain field for the 90° specimen in the slice that contains probe 1 superimposed with the corresponding slice of the laminography images. The location of the probe is indicated by a blue box for the first increment. Although considerable void growth is evidenced, no coalescence instances are observed. Intense strained bands develop between neighboring voids. The strained bands may span multiple voids. The invervoid ligament that contains probe 1 is larger than the ligaments that typically present strongly strained bands. This intervoid ligament shows a strained band in zigzag pattern and a round and more intense strain concentration zone in one corner of this zigzag pattern.

The strain field history in the slice containing probe 2 of the 90° specimen is shown in Figure 9. In this slice, intense strained bands are observed between neighboring voids, which are very close to each other as early as increment 10 (Figure 9(b)). Coalescence by internal necking occurs in some of these strained bands at increment 19 (Figure 9(c)) including the position of probe 2.

The history of the strain field of the 45° sample in the slice containing probe 1 is reported in

**Figure 8** Strain distribution in the slice of the computational domain that contains the center of probe 1 for the 90° specimen at increments (a) 1, (b) 10, and (c) 19. Position of probe 1 indicated by the blue box for increment 1. Strain map laid over the corresponding laminography section.

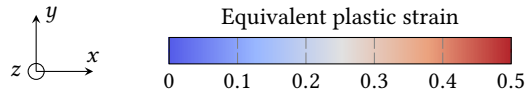
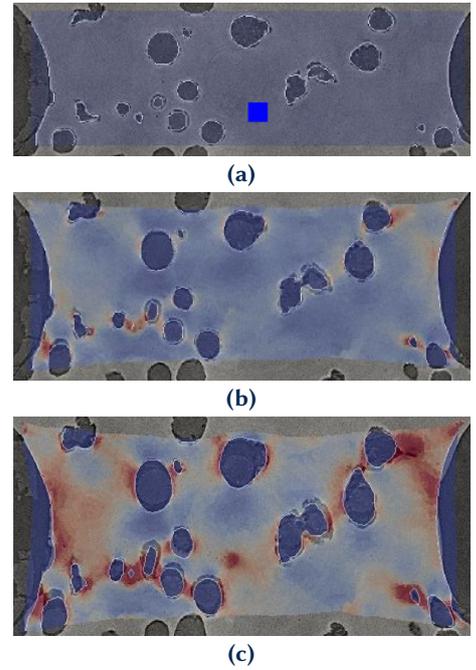

**Figure 9** Strain distribution in the slice of the computational domain that contains the center of probe 2 for the 90° specimen at increments (a) 1, (b) 10, and (c) 19. Oosition of probe 2 indicated by the red box for increment 1. Strain map laid over the corresponding laminography section.

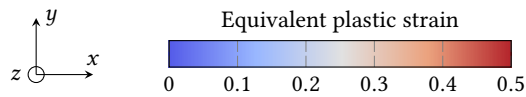
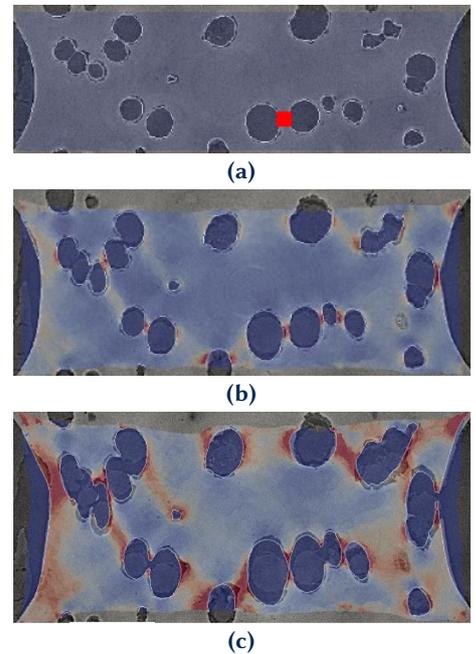

Figure 10. Only one instance of coalescence by internal necking is observed in a particularly small intervoid ligament. The strained band that covers the location of probe 1 is straight and diffuse.

**Figure 10** Strain distribution in the slice of the computational domain that contains the center of probe 1 for the 45° specimen at increments (a) 1, (b) 6, and (c) 11. Position of probe 1 indicated by the blue box for increment 1. Strain map laid over the corresponding laminography section.

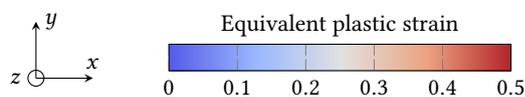
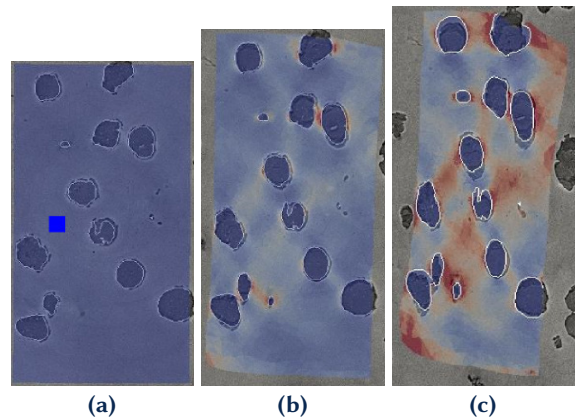

Figure 11 shows how the strain field evolves in the slice of the 45° specimen that contains probe 2. Increment 6, in Figure 11(b), exhibits a few intense strained bands that result in coalescence at increment 11, see Figure 11(c). Intense strained bands are observed at increment 11. The intervoid ligament where probe 2 is located does not present a distinct band.

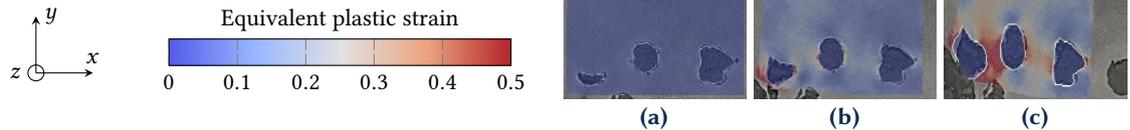

**Figure 11** Strain distribution in the slice of the computational domain that contains the center of probe 2 for the 45° specimen at increments (a) 1, (b) 6, and (c) 11. Position of probe 1 indicated by the blue box for increment 1. Strain map laid over the corresponding laminography section.

The positions of the probes were chosen in (Buljac et al. 2018a) to study, via DVC measurements, void coalescence through two different mechanisms, namely, void sheet coalescence (probe 1 of both specimens) and internal necking (probe 2 of both specimens). The authors found the equivalent strain to be a promising local indicator of coalescence and a strain threshold of approximately 40 % for the void sheet mechanism and 45 % for internal necking. However, if the observed instance of coalescence is taken into account for probe 2 of the 90° specimen, which corresponds to increment 11 in Figure 5(b), the observed precoalescence strain level corresponds to 20 %. Since coalescence occurred at a strain level considerably less than the threshold proposed in the mentioned work, the equivalent strain may not be a suitable coalescence indicator. In the present work, the choice of the equivalent plastic strain as a local indicator of void coalescence is revisited with the results of the FE simulations that provide an improved spatial resolution with respect to DVC measurements. If the equivalent strain is a sufficient and accurate coalescence indicator, coalescence should be observed wherever the threshold is exceeded. To test this hypothesis, precoalescence strain fields will be compared with the postcoalescence laminography images in the slices where the probes lie to confirm the occurrence of coalescence.

Figure 12 shows the precoalescence strain field in the slice where probe 1 is located at increment 19 (Figure 12(a)) and the corresponding postcoalescence slice at increment 20 (Figure 12(b)) of the 90° specimen. The slice reveals a crack that traverses most of the specimen section. The position of the crack follows the strained bands, the less intense part of which precisely corresponds to the ligament that contains probe 1. Two instances of highly strained intervoid ligaments that do not exhibit coalescence are observed and shown with green arrows in Figure 12(a).

The ligament that contains probe 1 presents the previously described strained band in zigzag pattern, which may be an indicator of interactions with microstructural heterogeneities in the depth direction. In fact, white contours to the left of the box confirm the presence of a void in approximately the same position of the probe hidden in the depth direction. Previous work has suggested that three-dimensional interactions such as this one may be at the origin of apparent void-sheet instances that are in fact a combination of two internal necking instances in a three-void cluster (Trejo Navas et al. 2018). This observation does not rule out the presence of pure void-sheet instances as the examination of fractured surfaces has provided hints of this mechanism (Buljac et al. 2018a). However, the confirmed presence of the third void hidden in the ligament in the depth direction cannot be ignored in this particular instance.

The precoalescence strain field for the 90° specimen at increment 10 (Figure 12(c)) and the corresponding postcoalescence slice at increment 11 (Figure 12(d)) are reported for the slice corresponding to probe 2. Six intervoid ligaments exhibit bands with high strain levels, two of which do not coalesce in the following increment. The size of the ligaments that do not undergo coalescence (denoted with a green arrow) does not vary with respect to the size of the ligaments that experience coalescence.

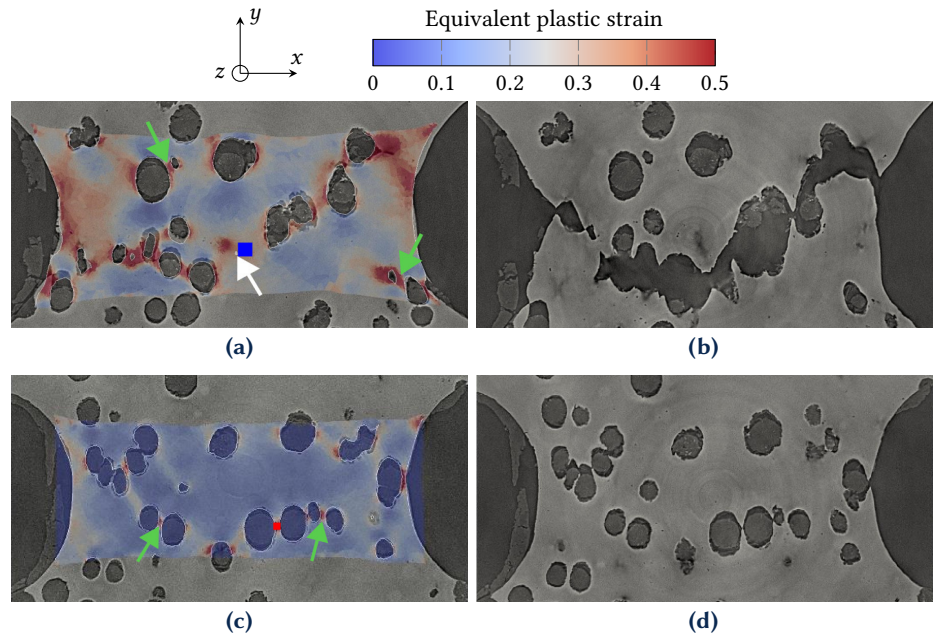

**Figure 12**    Sections of the 90° specimen: probe 1 (a) before coalescence (increment 19) and (b) after coalescence (increment 20); probe 2 (c) before coalescence (increment 10) and (d) after coalescence (increment 11). Positions of probes 1 and 2 indicated by the blue and red boxes, respectively. Occurrence of coalescence in the locations highlighted by the arrows discussed in the text.

The strain field prior to coalescence and the corresponding postcoalescence configuration for the 45° specimen are reported for the slice corresponding to probes 1 and 2 in Figure 13. Figure 13(b) evidences a crack at approximately 45° with respect to the loading (vertical) direction. The crack position does not correspond to the most strained parts of the domain and it crosses a ligament that did not belong to a strained band. A highly strained ligament that does not undergo coalescence is observed in this slice (marked with a green arrow). Similarly, in the slice corresponding to probe 2 (Figure 13(d)), the observed crack crosses ligaments that gave no indication of coalescence based on the strain levels and two highly strained ligaments do not lead to coalescence.

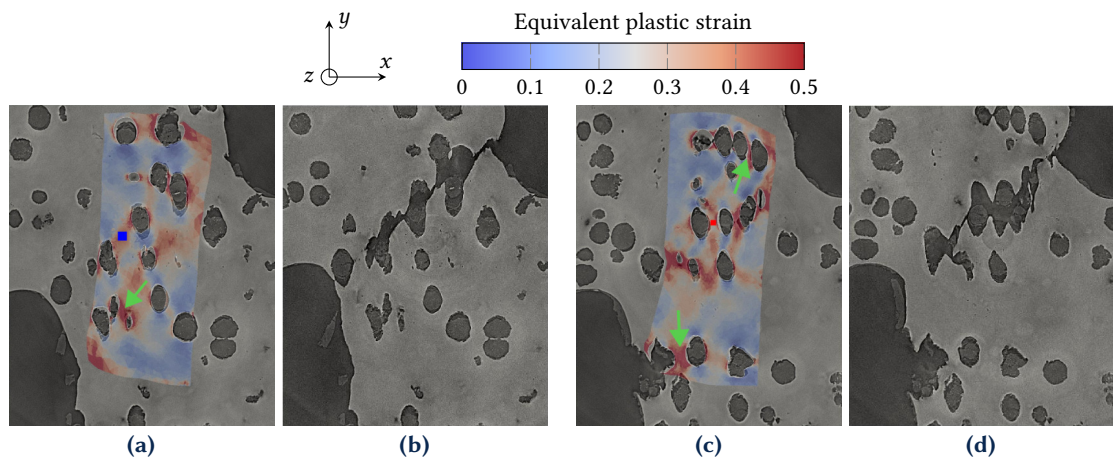

**Figure 13**    Sections of the 45° specimen: probe 1 (a) before coalescence (increment 10) and (b) after coalescence (increment 11); probe 2 (c) before coalescence (increment 10) and (d) after coalescence (increment 11). Positions of probes 1 and 2 indicated by the blue and red boxes, respectively. Occurrence of coalescence in the locations indicated by the arrows discussed in the text.

Although the equivalent strain may be an indicator of coalescence, such as in the slice containing probe 1 of the 90° specimen (Figure 12), the observation of these slices shows that highly strained instances that do not experience coalescence immediately after reaching high levels of strain, are not uncommon. A perspective to this work is to carry out more extensive quantitative analyses that could include, in addition to the equivalent strain, other strain measures and topological information such as initial intervoid distance.

Time discretization may be an intervening factor. Between two consecutive increments, the unobserved microstructural changes may induce variations in the strain distribution. This feature could explain why some apparently less strained ligaments seem to undergo coalescence before apparently more strained ligaments. Additional scans would be necessary to improve the temporal discretization and confirm this hypothesis. Using cubic probes of size of a DVC element were useful for validation purposes, but a more appropriate probe shape and size could be used to take advantage of the spatial resolution of FE simulations. Section 3.3 discusses this aspect.

## 3.3 Effect of probe size on FE strain evaluation

The local strain data presented in Sections 3.1 and 3.2 were calculated in cubic probes with the size of one DVC element in order to establish a direct comparison. When intervoid ligaments decrease, such DVC probes may become larger than the ligament and DVC strain results accuracy may be questionable. Since the FE simulations allow for more local strain evaluations, the effect of the probe size is studied in this section. First, a mesh sensitivity analysis is carried out to check that the results remain adequately mesh-independent when smaller probe sizes are employed. Then the effect of the probe size on local strain levels is assessed. The edge of the cubic probe will be scaled with a factor $f$. When $f = 1$, the probe size corresponds to that of the DVC element.

### 3.3.1 Mesh sensitivity analysis

A local mesh refinement was carried out around the probes to investigate mesh dependence and to correctly reproduce the desired cubic shape. This mesh refinement is depicted in Figure 14. In the refined volume, the isotropic metric $M$ was imposed the value $M = \text{diag}(1/(f_m h_{\min})^2)$, where $f_m$ is a coefficient that will be varied to produce different levels of local refinement and $h_{\min}$ is the smallest allowed mesh size ($h_{\min} = 0.25\,\mu\text{m}$). For the sake of brevity, the analysis is presented only for the 90° specimen. In that analysis, the probe size corresponds to an edge scaling factor $f$ of 0.25. This means that the volume of the probe corresponds to (1/64)-th the volume of a DVC element.

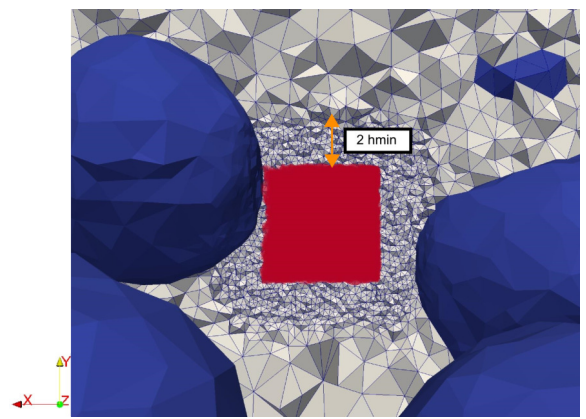

**Figure 14**  Probe 2 of the 45° specimen shown in red with the local mesh refinement around it for a slice of the domain. Voids shown in blue.

The selected values of $f_m$ are 1, 0.5 and 0.25. Figure 15 shows the mesh sensitivity results for probes 1 and 2. The maximum absolute difference between the datasets are also reported in the following figures comparing three or more datasets. The equivalent strain in probe 1 is not sensitive to the variations of $f_m$ in the chosen range, and a maximum difference of 0.02 is observed at the end of the simulation (for an equivalent strain level greater than 0.8). The equivalent strain in probe 2 is more sensitive. A maximum strain difference of 0.175 is obtained at the end of the simulation. The strain difference between the results with $f_m = 0.5$ and $f_m = 0.25$ is barely 0.03. A value of $f_m = 0.5$ is hence considered to produce mesh-independent results and will be used in the remainder of the study. The drop of strain in the final increment for probe 2 is caused by increased numerical diffusion of the probe volume due to an insufficiently fine mesh with $f_m = 1$.

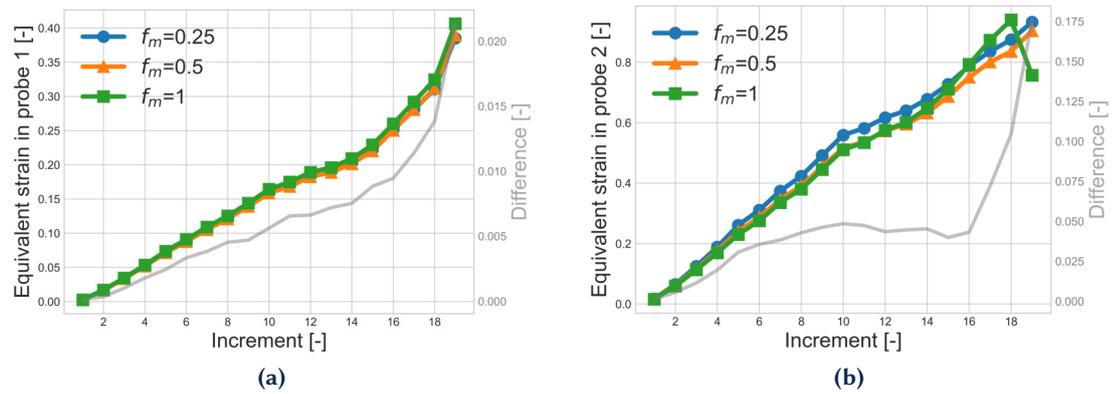

**Figure 15**	Mesh sensitivity of the equivalent strain of the 90° specimen: (a) probe 1, (b) probe 2.

### 3.3.2 Probe size sensitivity analysis

The sensitivity of the local strain levels to the probe size is assessed in the sequel. The reported strains correspond to the volume-average of the equivalent strain within the probe volume. Thus, if significant strain gradients exist, the strains are sensitive to the probe size. Three different values of the edge scaling factor $f$ are employed, i.e. 1, 0.5 and 0.25. In order to have at least two different probe sizes that are entirely accommodated in the intervoid ligament of probe 2, and for this probe only, the values of $f$ are 1, 0.4 and 0.25. This strain evaluation corresponds to the most local possible configuration with the resolution of the FE mesh (mesh size approximately equal to 0.125 µm), i.e., to the strain in the element that contains the coordinates of the center of the probe, and will be referred to as point probe.

Figure 16 provides three-dimensional views of the different probe sizes for the 90° and 45° specimens. In the case of probe 2 of the 90° specimen, only the probe sizes corresponding to $f = 0.4$ and $f = 0.25$ are shown in order to provide a clearer visualization. The size corresponding to $f = 1$ for this probe was shown in Figure 3(a).

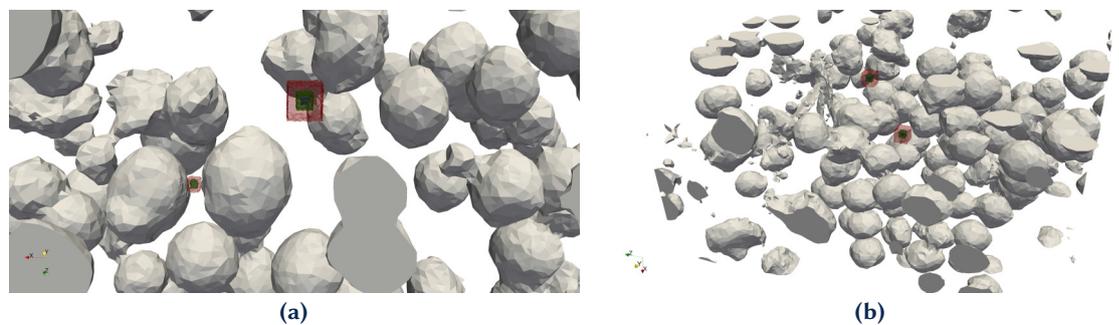

**Figure 16**	3D view of the microstructure with the corresponding probe sizes: (a) 90° specimen, (b) 45° specimen.

The sensitivity of the equivalent strain in probe 1 of the 90° specimen with respect to the probe size is shown in Figure 17(a). Little variation is observed for the four probe sizes. A maximum strain difference of 0.03 is obtained at the end of the simulation for a maximum level of 0.38. This is consistent with the observed strain fields (Figure 8). The probe is located in a diffuse zigzaged strain pattern without sharp gradients and, for this reason, the local measure of equivalent strain is not very sensitive to the probe size.

In Figure 17 the sensitivity of the equivalent strain of probe 2 of the 90° specimen is also studied with respect to the probe size. The probe size corresponding to $f = 1$ predicts strain levels that are considerably lower than those predicted by the other probe sizes. When $f = 1$, the probe is not small enough to fit in the intervoid ligament and the presence of the void phase in the probe volume alters the resulting strain. The remaining three sizes correspond to probes that fit in the intervoid ligament and lead to very similar results up to increment 14. In the subsequent increments, the prediction of the point probe strain starts to diverge from those with $f = 0.40$ and $f = 0.25$. At the end of the simulation, the former is lower than the latter with a strain difference of approximately 0.1. This difference results from sharp gradients to which the probe is submitted after flowing out of the intervoid ligament. Coalescence occurs at increment 11 for this ligament (Figure 5(b)). The results are satisfactory in the sense that the FE simulations can provide trustworthy strains from the beginning of the simulation and up to coalescence even in very narrow intervoid ligaments.

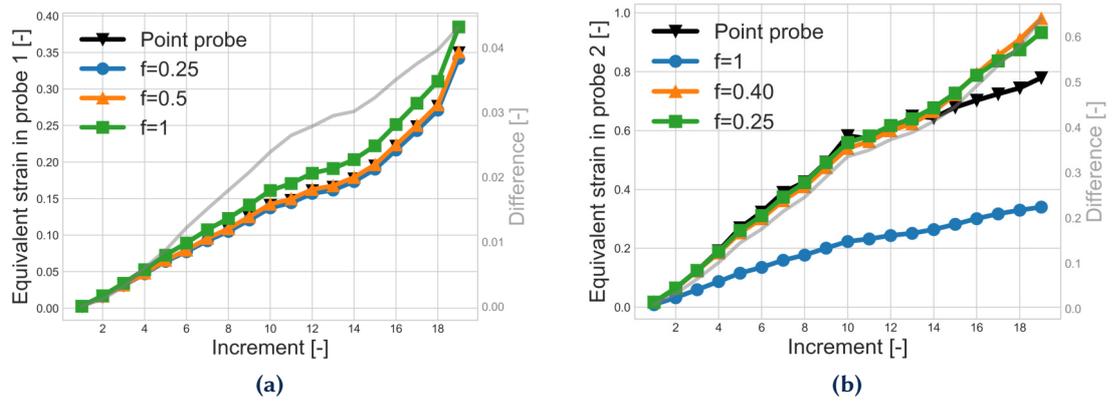

**Figure 17**  Probe size sensitivity of the equivalent strain of the 90° specimen: (a) probe 1, (b) probe 2.

The sensitivity to probe size of the strain measures associated with probes 1 and 2 of the 45° specimen are presented in Figure 18. For probe 1, the four strain levels are similar with a final maximum strain difference of 0.025 (for a maximum level of 0.36). When the probe size decreases (from $f = 1$ to $f = 0.25$), the strain increases monotonically. When the probe size further decreases, however, the strain level of the point probe lies between those corresponding to $f = 0.25$ and $f = 0.5$. For probe 2, a maximum strain difference of 0.03 is obtained and although the results of sizes associated with $f = 0.5$ and below are consistent during the whole simulation, a non monotonic trend is again observed with respect to probe size.

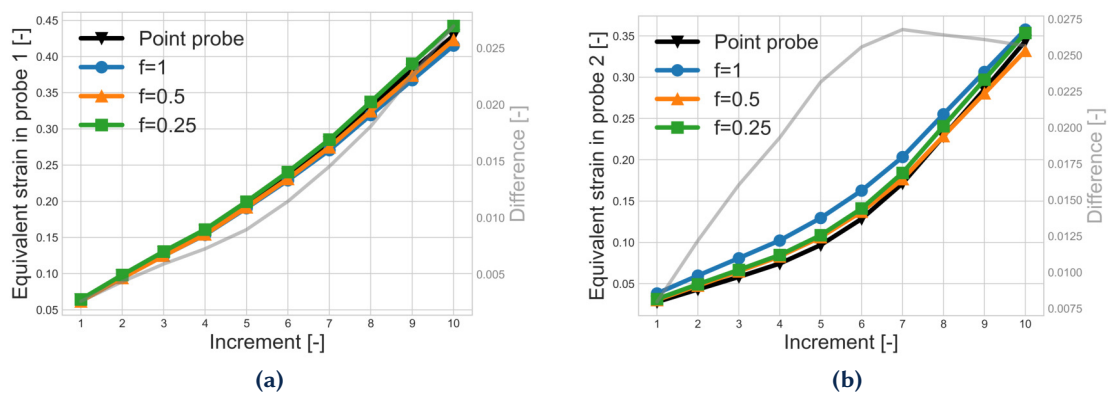

**Figure 18**  Probe size sensitivity of the equivalent strain of the 45° specimen: (a) probe 1, (b) probe 2.

To explore the non monotonic tendency found for the strains in probes 1 and 2 of the 45° specimen, the distribution of the equivalent strain in a slice containing probe 2 is shown in Figure 19 for increments 1, 4 and 8. The probes corresponding to $f = 1$, $f = 0.5$ and $f = 0.25$ are also depicted. The heterogeneity of the strain field in the intervoid ligament explains the non monotonic trend. For example, part of the largest probe is very close to the void-matrix interface, which corresponds to the most strained zones. This location increases the strain level in the probe beyond those observed in smaller probes. This result also illustrates how the strain evaluation in the probes depends on shape and orientation. This dependence is mitigated with the use of the point probe. For this reason, the point probe is considered the less ambiguous and yields the more local strains.

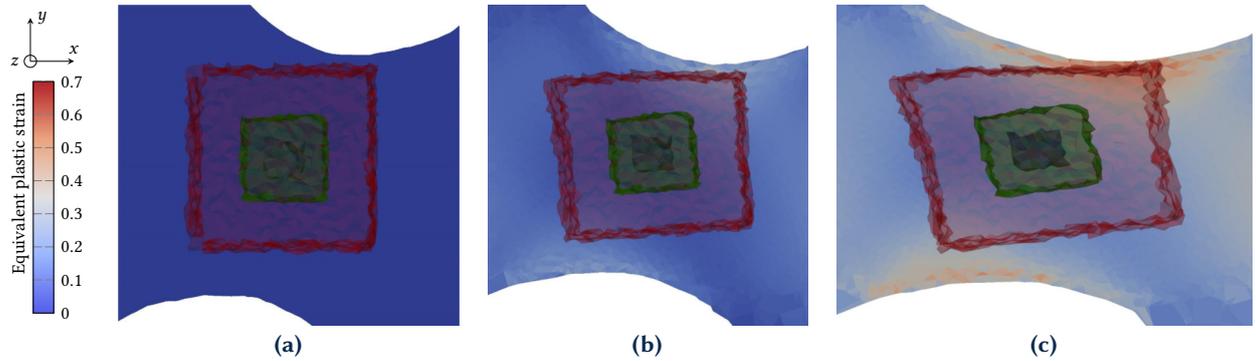

**Figure 19**    Distribution of the equivalent strain in a slice containing probe 2 of the 45° specimen for increments (a) 1, (b) 4 and (c) 8. Three probe sizes are shown.

## 4 Conclusions and perspectives

The main goal of this paper was to propose a hybrid experimental-numerical strategy that could give access to local strains at the onset of coalescence at intervoid ligaments. This is made possible here through the combined use of X-Ray *in-situ* laminography, DVC and FE computations.

A detailed comparison of DVC and FE strain evaluations was carried out herein based on images obtained by *in-situ* 3D synchroton imaging. The graphite nodules were explicitly meshed from the 3D images and were considered as voids. The boundary conditions were applied from DVC measurements of the 3D *in-situ* images. Macroscopic and three local estimates were reported for two different specimens and image segmentation-related uncertainties were taken into account.

The main conclusions are as follows:

- A very good agreement was observed for the macroscopic equivalent strain of both specimens as well as for the void precoalescence local strains of the specimen containing two machined holes oriented at 90° with respect to the loading direction.
- The rate of increase of the strain was well predicted for the local strain of the specimen containing two machined holes oriented at 45° with respect to the loading direction, but the absolute strain levels were not captured accurately. The FE strain estimates were, however, considered as satisfactory and suitable to be used in the study of void coalescence.
- The comparison between DVC measurements and FE predictions revealed that the image segmentation-related uncertainties were negligible with respect to the differences between the strain levels. This observation implies that the reported differences can be used to assess limitations of the model and numerical framework.
- When the specimen approached final fracture, an acceleration of the rate of strain was observed in the DVC results. The FE simulations failed to capture this acceleration. This difference is thought to be a deficiency of the hardening law for very large strain levels since it did not saturate or even soften. This shortcoming may be solved in future works by using a hardening law with saturation for large strain values or a constitutive law for porous plasticity or a coupled damage model (Lemaitre 1985). Accounting for damage in the matrix may be justified if nucleation at a second population of small particles is considered. A new parameter identification step at the microscale would be necessary.
- A sensitivity analysis of the local strain estimates with respect to the probe size was carried out. The strain levels were mostly sensitive in small intervoid ligaments. A point probe of the size of a FE element provided the most local information without dependence on probe shape and spatial orientation. This local information will be used in future investigations. The DVC analyses were less suitable than FE simulations because of their limited spatial resolution. The improved local precision of FE simulations highlight the importance of the combined SRCL-DVC-FE approach.
- To assess the suitability of the equivalent strain as a coalescence indicator, the strain fields were reported in detail for different sections of the three-dimensional simulations. The precoalescence strain fields were compared with the corresponding postcoalescence sections of the laminography images. The precoalescence equivalent strain could reliably be considered a coalescence indicator in several cases, but a non negligible number of highly strained instances that did not undergo coalescence was observed.

The results presented in this work call into question the possibility of using the precoalescence strain alone as a local coalescence criterion. The FE simulations allow the stress state to be used in such analyses. A more extensive quantitative study considering topological data as well as strain and stress evaluations is a perspective to this work.

## References

Bandstra, J. and D. Koss (2001). Modeling the ductile fracture process of void coalescence by void-sheet formation. *Materials Science and Engineering: A* 319:490–495. [DOI].

Bay, B., T. Smith, D. Fyhrie, and M. Saad (1999). Digital volume correlation: three-dimensional strain mapping using X-ray tomography. *Experimental mechanics* 39(3):217–226. [DOI].

Brezzi, F., D. Boffi, L. Demkowicz, R. Durán, R. Falk, and M. Fortin (2008). *Mixed finite elements, compatibility conditions, and applications*. Vol. 1939. Lecture Notes in Mathematics. Springer. [DOI].

Buljac, A., L. Helfen, F. Hild, and T. Morgeneyer (2018a). Effect of void arrangement on ductile damage mechanisms in nodular graphite cast iron: In situ 3D measurements. *Engineering Fracture Mechanics* 192:242–261. [DOI], [HAL].

Buljac, A., M. Shakoor, J. Neggers, M. Bernacki, P.-O. Bouchard, L. Helfen, T. Morgeneyer, and F. Hild (2017). Numerical validation framework for micromechanical simulations based on synchrotron 3D imaging. *Computational Mechanics* 59(3):419–441. [DOI], [HAL].

Buljac, A., T. Taillandier-Thomas, L. Helfen, T. Morgeneyer, and F. Hild (2018b). Evaluation of measurement uncertainties of digital volume correlation applied to laminography data. *The Journal of Strain Analysis for Engineering Design* 53(2):49–65. [DOI], [HAL].

Buljac, A., T. Taillandier-Thomas, T. Morgeneyer, L. Helfen, S. Roux, and F. Hild (2016). Slant strained band development during flat to slant crack transition in AA 2198 T8 sheet: in situ 3D measurements. *International Journal of Fracture* 200(1-2):49–62. [DOI], [HAL].

Buljac, A., V. M. Trejo Navas, M. Shakoor, A. Bouterf, J. Neggers, M. Bernacki, P.-O. Bouchard, T. Morgeneyer, and F. Hild (2018c). On the calibration of elastoplastic parameters at the microscale via X-ray microtomography and digital volume correlation for the simulation of ductile damage. *European Journal of Mechanics-A/Solids* 72:287–297. [DOI], [HAL].

Digonnet, H., L. Silva, and T. Coupez (2007). Cimlib: a fully parallel application for numerical simulations based on components assembly. *9th International Conference on Numerical Methods in Industrial Forming Processes* (Porto (Portugal), June 17, 2007–June 21, 2007). Vol. 908. AIP Conference Proceedings, pp 269–274. [DOI], [HAL].

El Khaoulani, R. and P.-O. Bouchard (2013). Efficient numerical integration of an elastic-plastic damage law within a mixed velocity-pressure formulation. *Mathematics and Computers in Simulation* 94:145–158. [DOI], [HAL].

Fischer, G., J. Nellesen, N. Anar, K. Ehrig, H. Riesemeier, and W. Tillmann (2013). 3D analysis of micro-deformation in VHCF-loaded nodular cast iron by μCT. *Materials Science and Engineering: A* 577:202–209. [DOI].

Gammage, J., D. Wilkinson, D. Embury, and E. Maire (2005). Damage studies in heterogeneous aluminium alloys using X-ray tomography. *Philosophical Magazine* 85(26-27):3191–3206. [DOI], [HAL].

Helfen, L., T. Baumbach, P. Mikulik, D. Kiel, P. Pernot, P. Cloetens, and J. Baruchel (2005). High-resolution three-dimensional imaging of flat objects by synchrotron-radiation computed laminography. *Applied Physics Letters* 86(7):071915. [DOI], [HAL].

Huang, Y. (1991). Accurate dilatation rates for spherical voids in triaxial stress fields. *Journal of Applied Mechanics* 58:1084. [DOI], [HAL].

Hütter, G., L. Zybell, and M. Kuna (2015). Micromechanisms of fracture in nodular cast iron: From experimental findings towards modeling strategies-A review. *Engineering Fracture Mechanics* 144:118–141. [DOI], [HAL].

Kasvayee, K. A., K. Salomonsson, E. Ghassemali, and A. E. Jarfors (2016). Microstructural strain distribution in ductile iron; comparison between finite element simulation and digital image correlation measurements. *Materials Science and Engineering: A* 655:27–35. [DOI].

Koplik, J. and A. Needleman (1988). Void growth and coalescence in porous plastic solids. *International Journal of Solids and Structures* 24(8):835–853. [DOI].

Kuna, M. and D. Sun (1996). Three-dimensional cell model analyses of void growth in ductile materials. *International Journal of Fracture* 81(3):235–258. [DOI].

Lemaitre, J. (1985). A continuous damage mechanics model for ductile fracture. *Journal of engineering materials and technology* 107(1):83–89. [DOI], [HAL].

Ludwik, P. (1909). *Elemente der Technologischen Mechanik.* [DOI].

Papasidero, J., V. Doquet, and D. Mohr (2015). Ductile fracture of aluminum 2024-T351 under proportional and non-proportional multi-axial loading: Bao-Wierzbicki results revisited. *International Journal of Solids and Structures* 69:459–474. [DOI], [HAL].

Pineau, A., A. Benzerga, and T. Pardoen (2016). Failure of metals I: Brittle and ductile fracture. *Acta Materialia* 107:424–483. [DOI], [HAL].

Quan, D.-L., T. Toulorge, E. Marchandise, J.-F. Remacle, and G. Bricteux (2014). Anisotropic mesh adaptation with optimal convergence for finite elements using embedded geometries. *Computer Methods in Applied Mechanics and Engineering* 268:65–81. [DOI], [HAL].

Roux, S., F. Hild, P. Viot, and D. Bernard (2008). Three dimensional image correlation from X-Ray computed tomography of solid foam. *Composites Part A: Applied Science and Manufacturing* 39(8):1253–1265. [DOI], [HAL].

Salomonsson, K. and J. Olofsson (2017). Analysis of Localized Plastic Strain in Heterogeneous Cast Iron Microstructures Using 3D Finite Element Simulations. *4th World Congress on Integrated Computational Materials Engineering* (Ypsilanti, USA, May 21, 2017–May 25, 2017), pp 217–225. [DOI], [HAL].

Scheyvaerts, F., T. Pardoen, and P. Onck (2010). A new model for void coalescence by internal necking. *International Journal of Damage Mechanics* 19(1):95–126. [DOI], [HAL].

Seleš, K., A. Jurčević, Z. Tonković, and J. Sorić (2019). Crack propagation prediction in heterogeneous microstructure using an efficient phase-field algorithm. *Theoretical and Applied Fracture Mechanics* 100:289–297. [DOI].

Shakoor, M., M. Bernacki, and P.-O. Bouchard (2015). A new body-fitted immersed volume method for the modeling of ductile fracture at the microscale: Analysis of void clusters and stress state effects on coalescence. *Engineering Fracture Mechanics* 147:398–417. [DOI], [HAL].

Shakoor, M., M. Bernacki, and P.-O. Bouchard (2017a). Ductile fracture of a metal matrix composite studied using 3D numerical modeling of void nucleation and coalescence. *Engineering Fracture Mechanics* 189:110–132. [DOI], [HAL].

Shakoor, M., P.-O. Bouchard, and M. Bernacki (2017b). An adaptive level-set method with enhanced volume conservation for simulations in multiphase domains. *International Journal for Numerical Methods in Engineering* 109(4):555–576. [DOI], [HAL].

Shakoor, M., A. Buljac, J. Neggers, F. Hild, T. Morgeneyer, L. Helfen, M. Bernacki, and P.-O. Bouchard (2017c). On the choice of boundary conditions for micromechanical simulations based on 3D imaging. *International Journal of Solids and Structures* 112:83–96. [DOI], [HAL].

Shen, Y., T. Morgeneyer, J. Garnier, L. Allais, L. Helfen, and J. Crépin (2013). Three-dimensional quantitative in situ study of crack initiation and propagation in AA6061 aluminum alloy sheets via synchrotron laminography and finite-element simulations. *Acta Materialia* 61(7):2571–2582. [DOI].

Tekkaya, A, P.-O. Bouchard, S. Bruschi, and C. Tasan (2020). Damage in metal forming. *CIRP Annals* 69:600–623. [DOI], [HAL].

Tomičević, Z., J. Kodvanj, and F. Hild (2016). Characterization of the nonlinear behavior of nodular graphite cast iron via inverse identification. Analysis of uniaxial tests. *European Journal of Mechanics-A/Solids* 59:140–154. [DOI], [HAL].

Trejo Navas, V. M., M. Bernacki, and P.-O. Bouchard (2018). Void growth and coalescence in a three-dimensional non-periodic void cluster. *International Journal of Solids and Structures* 139:65–78. [DOI], [HAL].

Trejo Navas, V. M., A. Buljac, F. Hild, T. Morgeneyer, L. Helfen, M. Bernacki, and P.-O. Bouchard

(2019). A comparative study of image segmentation methods for micromechanical simulations of ductile damage. *Computational Materials Science* 159:43–65. [DOI], [HAL].

Voce, E. (1955). A practical strain hardening function. *Metallurgia* 51:219–226.

Weck, A. and D. Wilkinson (2008). Experimental investigation of void coalescence in metallic sheets containing laser drilled holes. *Acta Materialia* 56(8):1774–1784. [DOI].

Weck, A., D. Wilkinson, and E. Maire (2008). Observation of void nucleation, growth and coalescence in a model metal matrix composite using X-ray tomography. *Materials Science and Engineering: A* 488(1):435–445. [DOI], [HAL].

**Authors' contributions**   X-Ray laminography was carried out by Ante Buljac under the supervision of Thilo Morgeneyer. DVC analyses were done by Ante Buljac under the supervision of François Hild. FE computations were performed by Victor Trejo Navas under the supervision of Marc Bernacki and Pierre-Olivier Bouchard. Victor Trejo Navas wrote the initial draft of the paper and all the authors discussed the results and wrote the final version.

**Supplementary Material**   None.

**Acknowledgements**   This work was performed within the COMINSIDE project funded by the French Agence Nationale de la Recherche (ANR-14-CE07-0034-02 grant). We thank ESRF for allowing beamtime me1366 at iD15A. Lukas Helfen from KIT is acknowledged for help with synchrotron imaging. TU Freiberg is thanked for materials supply.

**Ethics approval and consent to participate**   Not applicable.

**Consent for publication**   Not applicable.

**Competing interests**   The authors declare that they have no competing interests.

**Journal's Note**   JTCAM remains neutral with regard to the content of the publication and institutional affiliations.